\documentclass[12pt,letterpaper,a4paper]{article}
\usepackage[includeheadfoot,
            marginratio={1:1,2:3},
            width=450pt,
            height=700pt,]{geometry}
\usepackage{amsmath}
\usepackage{amsfonts}
\usepackage{amssymb}
\usepackage{graphicx}
\usepackage{cancel}

\usepackage{empheq}
\usepackage{color}
\usepackage{hyperref}

\usepackage{float}
\restylefloat{table}
\restylefloat{figure}

\newcommand{\nc}{\newcommand}
\nc{\beq}{\begin{equation}}
\nc{\eeq}{\end{equation}}
\nc{\bea}{\begin{eqnarray}}
\nc{\eea}{\end{eqnarray}}
\def\IZ{\mathbb{Z}}

\def\ov{\overline}

\begin{document}

\vspace{1.5cm}
\begin{center}
{\LARGE
Dimensional oxidation and modular \vskip0.15cm completion of  non-geometric type IIB action}
\vspace{0.4cm}
\end{center}

\vspace{0.35cm}
\begin{center}
 Xin Gao$^\dag$\footnote{Email: xingao@vt.edu}  and Pramod Shukla$^\ddag$\footnote{Email: pkshukla@to.infn.it}
\end{center}

\vspace{0.1cm}
\begin{center}
{
$^{\dag}$Department of Physics, Robeson Hall,  Virginia Tech,
\vskip0.02cm
850 West Campus Drive, Blacksburg, VA 24061, USA \\
\vskip0.4cm
$^{\ddag}$Universit\'a di Torino, Dipartimento di Fisica and I.N.F.N.-sezione di Torino
\vskip0.02cm
Via P. Giuria 1, I-10125 Torino, Italy
}
\end{center}

\vspace{1cm}
\begin{abstract}
Utilizing a setup of type IIB superstring theory compactified on an orientifold of ${\mathbb T}^6/{\left({\mathbb Z}_2 \times {\mathbb Z}_2\right)}$, we propose a modular invariant dimensional oxidation of the four-dimensional scalar potential. In the oxidized ten-dimensional supergravity action, the standard NS-NS and RR three form fluxes ($H$-, $F$-) as well as the non-geometric fluxes ($Q$-, $P$-) are found to nicely rearrange themselves to form generalized flux-combinations. 
As an application towards moduli stabilization, using the same S-duality invariant scalar potential, we examine the recently proposed No-Go theorem \cite{Blumenhagen:2014nba} about creating a mass-hierarchy between universal-axion and the dilaton relevant for axionic-inflation. Considering a two-field dynamics of universal axion and dilator while assuming the other moduli/axions being stabilized, we find a part of the No-Go arguments to be quite robust even with the inclusion of non-geometric ($Q$-, $P$-) fluxes.
\end{abstract}

\clearpage

\tableofcontents

\section{Introduction}
\label{sec:intro}

Orientifolds of Type II superstring theories admit generalized fluxes via a successive application of T-duality on the three form $H$-flux. The same results in a chain of geometric and non-geometric fluxes as
\bea
\label{eq:Tdual}
& & H_{ijk} \longrightarrow \omega_{jk}{}^i  \longrightarrow Q_k{}^{ij}  \longrightarrow R^{ijk},
\eea
and have led to impetus progress in constructing string solutions in connection with the gauged supergravities in recent years \cite{Kachru:2002sk,Hellerman:2002ax,Dabholkar:2002sy,Hull:2004in,Derendinger:2004jn, Derendinger:2005ph, Shelton:2005cf, Dall'Agata:2009gv, Aldazabal:2011yz, Aldazabal:2011nj,Geissbuhler:2011mx,Grana:2012rr,Dibitetto:2012rk, Andriot:2012wx, Andriot:2012an, Andriot:2011uh,Blumenhagen:2013hva,Andriot:2013xca,Andriot:2014qla,Blair:2014zba}. Generically, all of  such fluxes appear as parameters in the four dimensional effective potential and hence can develop a suitable scalar potential for the purpose of moduli stabilization which has been among the central aspects towards constructing realistic string models. For this goal, it is always preferred to have compactification backgrounds of much more rich structure and as much ingredients as possible because the same can induce new possibilities to facilitate the demands of (semi-)realistic model building. On these lines, the application of non-geometric fluxes towards moduli stabilization and cosmological model building aspects have attracted great amount of interest \cite{deCarlos:2009qm, Danielsson:2012by, Blaback:2013ht, Damian:2013dq, Damian:2013dwa, Hassler:2014mla} in recent time.

String fluxes are closely related to the possible gaugings in the gauged supergravity \cite{ Derendinger:2004jn, Derendinger:2005ph, Shelton:2005cf, Aldazabal:2006up, Dall'Agata:2009gv, Aldazabal:2011yz, Aldazabal:2011nj,Geissbuhler:2011mx,Grana:2012rr,Dibitetto:2012rk, Villadoro:2005cu} and it is remarkable that the four dimensional effective potentials could be studied (without having a full understanding of their ten-dimensional origin) via merely knowing the forms of K\"ahler and super-potentials \cite{Danielsson:2012by, Blaback:2013ht, Damian:2013dq, Damian:2013dwa, Blumenhagen:2013hva, Villadoro:2005cu, Robbins:2007yv, Ihl:2007ah}. Being simpler and well understood in nature, the Type II toroidal orinetifolds provide a promising toolkit to begin with while looking at new aspects, and so is the case with investigating the effects of non-geometric fluxes. Further, unlike the case with Calabi Yau compactifications, the explicit and analytic form of metric being known for the toroidal compactification backgrounds make such backgrounds automatically the favorable ones for performing explicit computations and studying the deeper insights of non-geometric aspects; for example the knowledge of metric has helped in knowing the ten-dimensional origin of the geometric flux dependent \cite{Villadoro:2005cu} as well as the non-geometric flux dependent potentials \cite{Blumenhagen:2013hva}.
In our previous work \cite{Blumenhagen:2013hva},  we have performed a close investigation
of the effects of T-duality motivated fluxes via considering their presence in the induced four-dimensional superpotential as proposed in \cite{Shelton:2005cf, Aldazabal:2011nj}.
We determined the most general form of the $H$, $F$, $Q$, $R$-fluxes in terms of the generalised metric and derived the Bianchi identities among these fluxes.
On a simple toroidal orientifold of type IIA and its T-dual type IIB model with all T-dual invariant geometric and non-geometric NS-NS
and R-R fluxes turned-on, we have computed the induced scalar potential from the four-dimensional superpotential and subsequently we have oxidized the various pieces into an underlying ten-dimensional supergravity action  \cite{Blumenhagen:2013hva}.
We found that, both in the NS-NS and in the R-R sector, the resulting
oxidized ten-dimensional action is compatible with the flux formulation of the Double Field Theory action \cite{Aldazabal:2011nj,Geissbuhler:2011mx}.


The connection between a string compactification and the gauged supergravity effective theory mentioned so far is not the full story \cite{Aldazabal:2008zza,Font:2008vd,Guarino:2008ik} for both the type II superstring theories. In a setup of type IIB superstring theory compactified on ${\mathbb T}^6/{\left({\mathbb Z}_2 \times {\mathbb Z}_2\right)}$, it was argued that the additional fluxes are needed to ensure S-duality invariance of underlying low energy type IIB supergravity. The resulting modular completed fluxes can be arranged into spinor representations of $SL(2,{\mathbb Z})^7$, and can be described globally via a non-geometric compactification of F-theory when there is a geometric local description in terms of ten-dimensional supergravity \cite{Aldazabal:2008zza}. The Jacobi identities of the flux algebra then lead to the general form of the Bianchi identites in F-theory compactifications. The compactification manifold with $T$- and $S$-duality appears to be an $U$-fold \cite{Hull:2004in, Aldazabal:2008zza, Kumar:1996zx, Hull:2003kr} where local patches are glued
by performing $T$- and $S$-duality transformations.
As a result, a generalization of our previous work \cite{Blumenhagen:2013hva} to include the S-dual version of  the non-geometric Q-flux, called P-flux, is necessary.
It is expected to provide a direct connection between the four-dimensional superpotential and the stringy aspects of the original  T- and S-duality
invariant ten-dimensional supergravity.

Recently, axionic-inflation has received a lot of interest due to the
possible detection of primordial gravitational waves claimed by the BICEP2 collaboration \cite{Ade:2014xna}. The recent result of PLANCK \cite{Ade:2015lrj} implies that the value of tensor-to-scalar ratio $(r)$ around 0.2 (as claimed by BICEP2) can be explained by the foreground dust. Nevertheless, because of having an upper bound as $r< 0.11$, constructing models to realize non-trivially large values of $r$ are compatible  as well as desired from the point of view of the possible future detection of gravitational waves. In the context of axion driven inflationary models developed in Type IIB/F-theory compactifciation, many proposals have emerged \cite{Hassler:2014mla,Grimm:2007hs,Blumenhagen:2014gta,Grimm:2014vva,Marchesano:2014mla,Arends:2014qca,Long:2014dta,Gao:2014uha,Ben-Dayan:2014lca,Garcia-Etxebarria:2014wla} in the recent times. In the original axion-monodromy inflation \cite{McAllister:2008hb, Flauger:2009ab}, the involutively odd $C_2$ axions have been proposed as being the inflaton candidate. The specific Calabi-Yau orientifolds which could support such odd axions along with their (F-term) moduli stabilization aspects have been studied in \cite{Blumenhagen:2008zz, Gao:2013pra,Grimm:2011dj,Gao:2013rra}. Regarding one of the recent axionic inflation models, a No-Go theorem has been proposed \cite{Blumenhagen:2014nba} as a challenge of creating a mass-hierarchy between universal axion and dilaton in type IIB orientifold compactifciation. The same has been of interest for constructing axion-monodromy inflationary model involving the universal axion \cite{Blumenhagen:2014gta,Gao:2014uha}. Equipped with the modular completed fluxes, we examine the original No-Go theorem.  We find that despite of relatively much richer structure for universal-axion/dilaton dependences of the full potential, quite surprisingly, the No-Go statement still holds in a two field analysis, and thus showing its robustness. 

This paper is organized as follows: in section \ref{sec:orientifold}, we present the basic set-up of the type IIB on $ {\mathbb T}^6 / \left(\mathbb Z_2\times \mathbb Z_2\right)$ orientifold and the general fluxes allowed to write out a generic form of superpotential involving the two NS-NS fluxes ($H, Q$) and their respective S-dual ($F, P$) fluxes. In section \ref{sec:Rearrangement}, a detailed study of the four-dimensional scalar potential induced by the flux superpotential is performed which takes us to propose a dimensional oxidation into the underlying ten-dimensional action in section {\ref{sec:S-dual}. The form of Chern-Simons terms reproducing the respective 3-brane and 7-brane tadpoles are also consistently invoked while considering the $SL(2,\IZ)$ invariance. Next, as an application to the potential we derived, in section \ref{sec:no-go}, we examine the role of non-geometric fluxes, specially S-dual P-fluxes, which they could play in the context of the No-go theorem mentioned in \cite{Blumenhagen:2014nba}. In the end, we summarize the results followed by two short appendices (\ref{appendix_0}) and (\ref{sec:D-term}) detailing some intermediate steps and the strategy followed for invoking the flux combinations needed for oxidation purpose.


\section{Type IIB on $ {\mathbb T}^6 / \left(\mathbb Z_2\times \mathbb Z_2\right)$ orientifold and fluxes}
\label{sec:orientifold}

Following the notations of \cite{Blumenhagen:2013hva}, let us briefly revisit the relevant features of a setup within type IIB superstring theory compactified on  $ {\mathbb T}^6 / \left(\mathbb Z_2\times \mathbb Z_2\right)$ with the
two $\mathbb Z_2$ actions being defined as
\bea
\label{thetaactions}
& & \theta:(z^1,z^2,z^3)\to (-z^1,-z^2,z^3)\\
 & &   \ov\theta:(z^1,z^2,z^3)\to (z^1,-z^2,-z^3)\, . \nonumber
 \eea
Further, the orientifold action is: ${\cal O} \equiv \Omega\,  I_6 \, (-1)^{F_L}$ where $\Omega$ is the worldsheet parity, $F_L$ is left-fermion number while the holomorphic involution $I_6$ being defined as
\bea
\label{eq:orientifold}
& & I_6 : (z^1,z^2,z^3)\rightarrow (-z^1,-z^2,-z^3)\,,
\eea
resulting in a setup with the presence of $O3/O7$-plane. The complex coordinates $z_i$'s on $T^6=T^2\times T^2\times T^2$ are defined as
\bea
z^1=x^1+ i U_1  x^2, ~ z^2=x^3+ i U_2 x^4,~ z^3=x^5+ i U_3 x^6 ,
\eea
where the three complex structure moduli $U_i$'s can be written as
$U_i= u_i + i\, v_i,\,\,i=1,2,3$. 

Now choosing the following basis of closed three-forms
\bea
\label{formbasis}
       \alpha_0&=dx^1\wedge dx^3\wedge dx^5\,, ~
       \beta^0=dx^2\wedge dx^4\wedge dx^6\, ,\nonumber\\
       \alpha_1&=dx^1\wedge dx^4\wedge dx^6\, , ~
       \beta^1=dx^2\wedge dx^3\wedge dx^5\, ,\\
       \alpha_2&=dx^2\wedge dx^3\wedge dx^6\, , ~
       \beta^2=dx^1\wedge dx^4\wedge dx^5\, ,\nonumber\\
      \alpha_3&=dx^2\wedge dx^4\wedge dx^5\, ,~
       \beta^3=dx^1\wedge dx^3\wedge dx^6\,  \nonumber
\eea
satisfying $\int \alpha_I\wedge \beta^J=-\delta_I{}^J$, the
holomorphic three-form $\Omega_3=dz^1\wedge dz^2\wedge dz^3$ can be expanded as
\bea
& & \Omega_3=  \alpha_0 + i \, (U_1 \beta^1 + U_2 \beta^2 + U_3 \beta^3)-i U_1
U_2 U_3 \beta^0 \nonumber\\
& & \hskip1.5cm -U_2 U_3 \alpha_1- U_1 U_3 \alpha_2 - U_1 U_2 \alpha_3\, .
\eea
The additional chiral variable are axion-dilaton
\bea
& & S= \, e^{-\phi}-i \, C_{0}\, .
\eea
and the K\"ahler moduli, generically being encoded in  the complexified  four-cycle volumes given as
\bea
J^c= \frac{1}{2} e^{-\phi} J \wedge J +i \, C^{(4)}\; .
\eea
In our case, these moduli are
\bea
T_1 = \tau_1+ i \, C^{(4)}_{3456}, ~ ~T_2 =\tau_2+ i \, C^{(4)}_{1256} , ~ ~T_3 = \tau_3+ i \, C^{(4)}_{1234},
\eea
where the real parts can be expressed in terms of the two-cycle volumes $t_i$
as,
$
\tau_1 = e^{-\phi} \, t_2 \, t_3, \,\,\tau_2 = e^{-\phi} \, t_3 \, t_1, \, \,
\,\,\tau_3 = e^{-\phi}  \, t_1 \, t_2.
$
We also need to express the two-cycle volumes $t_i$ in terms of the
four-cycles volumes $\tau_i$ as,
\bea
\label{twoinfour}
& & t_1=\sqrt{\tau_2\,  \tau_3\over \tau_1\, s}\,, ~~  t_2=\sqrt{\tau_1\,  \tau_3\over \tau_2\, s}\,, ~~ t_3=\sqrt{\tau_1\,  \tau_2\over \tau_3\, s}\,
\eea
with $s={\rm Re}(S)$.
Now, the non-vanishing components of the metric
in string frame are
\bea
\label{eq:gMN}
    g_{MN}={\rm blockdiag}\Big({e^{\phi\over 2}\over {\sqrt{\tau_1\, \tau_2
           \, \tau_3}}} \, \, \tilde g_{\mu\nu}, \, \, g_{ij}\Big) \, .
\eea
Further, the string frame internal metric $g_{ij}$
is also block-diagonal and
has the following non-vanishing components,
\bea
& & \hskip-0.8cm g_{11}=\frac{t_1}{u_1}\,,   ~ g_{12}=-\frac{t_1 v_1}{u_1} =g_{21}\, , ~
g_{22}=\frac{t_1(u_1^2+v_1^2)}{u_1}\, ,\nonumber\\
& & \hskip-0.8cm g_{33}=\frac{t_2}{u_2}\, ,~    g_{34}=-\frac{t_2 v_2}{u_2}=g_{43}\, , ~
g_{44}=\frac{t_2(u_2^2+v_2^2)}{u_2}\, ,\\
& & \hskip-0.8cm g_{55}=\frac{t_3}{u_3}\,, ~    g_{56}=-\frac{t_3 v_3}{u_3}=g_{65}\,, ~
g_{66}=\frac{t_3(u_3^2+v_3^2)}{u_3}\, .\nonumber
\eea
These internal metric components can be written out in more a suitable form, to be utilized later, by using the four cycle volumes $\tau_i$'s and the same is given as under,
\bea
& & \hskip-0.8cm g_{11}=\frac{\sqrt{\tau_2} \,  \sqrt{\tau_3}}{\sqrt{s} \, u_1 \, \sqrt{\tau_1}}\,,   ~ g_{12}=-\frac{v_1 \, \sqrt{\tau_2} \,  \sqrt{\tau_3}}{\sqrt{s} \, u_1 \, \sqrt{\tau_1}} =g_{21}\, , ~
g_{22}= \frac{\left(u_1^2 + v_1^2\right)\sqrt{\tau_2} \,  \sqrt{\tau_3}}{\sqrt{s} \, u_1 \, \sqrt{\tau_1}} ,\nonumber\\
& & \hskip-0.8cm g_{33}=\frac{\sqrt{\tau_1} \,  \sqrt{\tau_3}}{\sqrt{s} \, u_2 \, \sqrt{\tau_2}}\,,   ~ g_{34}=-\frac{v_2 \, \sqrt{\tau_1} \,  \sqrt{\tau_3}}{\sqrt{s} \, u_2 \, \sqrt{\tau_2}} =g_{43}\, , ~
g_{44}= \frac{\left(u_2^2 + v_2^2\right)\sqrt{\tau_1} \,  \sqrt{\tau_3}}{\sqrt{s} \, u_2 \, \sqrt{\tau_2}} ,\\
& & \hskip-0.8cm g_{55}=\frac{\sqrt{\tau_1} \,  \sqrt{\tau_2}}{\sqrt{s} \, u_3 \, \sqrt{\tau_3}}\,,   ~ g_{56}=-\frac{v_3 \, \sqrt{\tau_1} \,  \sqrt{\tau_2}}{\sqrt{s} \, u_3 \, \sqrt{\tau_3}} =g_{65}\, , ~
g_{66}= \frac{\left(u_3^2 + v_3^2\right)\sqrt{\tau_1} \,  \sqrt{\tau_2}}{\sqrt{s} \, u_3 \, \sqrt{\tau_3}} ,\nonumber
\eea
Since the background fluxes $\omega^i{}_{jk}$ and $R^{ijk}$ are odd under
the orientifold projection, the only invariant fluxes
are the following components of the three-forms $H_3$ and $F_{3}$ as under,
\bea
 & & \ov{H}:\quad {\ov H}_{135}\,,\  {\ov H}_{146}\, ,\  {\ov H}_{236}\, ,\  {\ov H}_{245}, {\ov H}_{246}\,, \ {\ov H}_{235}\, ,\ {\ov H}_{145}\, ,\ {\ov H}_{136}\, ,\\
 & &  \ov{F}:\quad  {\ov F}_{135}\,,\  {\ov F}_{146}\, ,\  {\ov F}_{236}\, ,\ {\ov F}_{245}\, , {\ov F}_{246}\,, \ {\ov F}_{235}\, ,\ {\ov F}_{145}\, ,\ {\ov F}_{136} \nonumber
\eea
and the  components of non-geometric $Q$ and P-fluxes, which can be collectively given as $\ov A\equiv Q$ or $P$:
\bea
& & \hskip-0.5cm \ov A: \, \, \, \ov A_1{}^{35}\,,\   \ov A_2{}^{45}\,,\   \ov
A_1{}^{46}\,,\   \ov A_2{}^{36}\, , \, \ov A_5{}^{13}\,,\   \ov A_6{}^{23}\,,\  \\
& & \hskip-0.5cm\hskip0.9cm \ov A_5{}^{24}\,,\   \ov A_6{}^{14}, \, \ov A_3{}^{51}\,,\   \ov A_4{}^{61}\,,\   \ov
A_3{}^{62}\,,\   \ov A_4{}^{52}\, ,\nonumber\\
& & \hskip-0.5cm \hskip0.9cm\ov A_2{}^{35}\,,\   \ov A_5{}^{23}\,,\   \ov
A_3{}^{52}\,,\   \ov A_2{}^{46} , \, \ov A_4{}^{51}\,,\   \ov A_1{}^{45}\,,\nonumber\\
& & \hskip-0.5cm \hskip0.9cm \ov A_5{}^{14}\,,\   \ov A_4{}^{62}\, , \, \ov A_6{}^{13}\,,\   \ov A_3{}^{61}\,,\   \ov
A_1{}^{36}\,,\   \ov A_6{}^{24}~. \nonumber
\eea
Now, the complete form of flux induced superpotential is given as \cite{Aldazabal:2006up},
\bea
\label{typeIIBW}
& & \hskip-0.5cm W =  \frac{1}{4} \int_X \left(\ov {F} - i \, S \, \ov{H}\right) \wedge \Omega_3 - \frac{i}{4}\int_{X} \left[\left(\ov{Q} \,- i \, S \, \ov{P}\right)\bullet
J^c\right]\wedge \Omega_3 ,
\eea
where the three-form of type $\ov{A}\bullet J^c={1\over 6} (\ov{A}\bullet J^c)_{ijk}\,
dx^i\wedge dx^j\wedge dx^k$ is defined as
\bea
    (\ov{A}\bullet J^c)_{ijk}={3\over 2}\, \ov{A}_{[\underline{i}}{}^{mn\,}
    J^c_{mn\underline{jk}]}\, \, \, \, \, \, \, \, \, {\rm for} \, \, \, A \in\{Q, P\}. \nonumber
\eea
Together with the   K\"ahler potential
\bea
\label{typeIIBK}
& & \hskip-0.5cm K = -\ln\left( \frac{S +\ov{S}}{2}\right) -\sum_{i=1,2,3}\ln\left( \frac{U_i +\ov{U_i}}{2}\right)- \, \sum_{i=1,2,3}\ln\left( \frac{T_i +\ov{T_i}}{2}\right)
\eea
it allows now to compute the F-term contribution of the effective four-dimensional scalar potential by utilizing the following standard relation
\bea
\label{VF}
V_{F}=e^{K}\Big(K^{i\bar\jmath}D_i W\, D_{\bar\jmath} \ov W-3
|W|^2\Big)\,.
\eea
As it is well reflected from the superpotential, the inclusion of dual P-flux provides a modular completion under the $SL(2, \IZ)$ transformation \cite{Aldazabal:2006up}:
\bea
\label{eq:SL2Z}
& & S\to \frac{kS-i\ell}{imS+n}\ , \quad kn-\ell m = 1\ , \quad k,\
\ell,\ m,\ n\in \mathbb{Z}\ ; \nonumber\\
& & \begin{pmatrix}F_3\\ H_3\end{pmatrix}\to\begin{pmatrix}k&\ell\\
m&n\end{pmatrix}\begin{pmatrix}F_3\\ H_3\end{pmatrix}, \, \, \, \, \, \begin{pmatrix}Q\\ P\end{pmatrix}\to\begin{pmatrix}k&\ell\\
m&n\end{pmatrix}\begin{pmatrix}Q\\ P\end{pmatrix}\ .
\eea
 Let us mention that for our example there are no two-forms
anti-invariant under the orientifold projection
so that  no $B_2$ and $C_2$ moduli are present.
The  $SL(2, \IZ)$ self-dual action for IIB will result in further transformation on the rest of the massless bosonic spectrum. To explicitly check the modular invariance of the four-dimensional scalar potential, we consider a simplified version of S-duality transformation given as $S \to {1}/{S}$, under which chiral variables, together with R-R field and various fluxes transform in the following manner,
\bea
\label{eq:S-duality}
& &   S \to \frac{1}{S}, \, \, T_{\alpha} \to {T_{\alpha}}, \, \, U_m \to {U_m}, \\
& & \hskip-0.0cm {H}_{ijk} \to {F}_{ijk}, \, \, {F}_{ijk} \to -{H}_{ijk}, {Q}^{ij}_{k} \to - {P}^{ij}_{k}, \, \, {P}^{ij}_{k}\to {Q}^{ij}_{k}. \nonumber
\eea
Here it should be noted that the Einstein-frame chiral coordinate $T_{\alpha}$ is invariant only in an orientifold with no odd axions, i.e. $h^{11}_-(X_6/{\cal O}) = 0$ \cite{Grimm:2007xm}. Under this S-duality, the superpotential and the K\"ahler potential (at the tree level) transform as:
\bea
\label{eq:S-duality1}
e^K \longrightarrow |S|^2 \, e^K\, , \, \, \, W \longrightarrow -\frac{i}{S} \, W.
\eea
which finally results in a S-duality invariant F-term potential $V_F$.

\section{Rearrangement of F-term scalar potential}
\label{sec:Rearrangement}
In this section we will present the full F-term scalar potential in the form of various ``suitable" pieces to be later utilized for the oxidation purpose in the next section.

Using the expressions of K\"ahler potential and superpotential given in eqns. (\ref{typeIIBW}-\ref{typeIIBK}), the full F-term scalar potential results in 9661 terms appearing in the form of quadratic-terms in four $H, F, Q$ and $P$-fluxes. Now, let us consider the following new flux-combinations which we have invoked after a very tedious terms-by-term investigation of the scalar potential,
\begin{subequations}
\label{eq:fluxOrbits}
\bea
\label{eq:fluxOrbits1}
&& {\cal H}_{ijk} =  {h}_{ijk}~,\quad\quad {\cal Q}^{ij}_{k} = {Q}^{ij}_{k} - C_0 \, ~{P}^{ij}_{k} ~, \\
&& {\cal F}_{ijk}= {f}_{ijk} - C_0 \, ~{h}_{ijk}~, \quad\quad {\cal P}^{ij}_{k} = {P}^{ij}_{k}~ ,\nonumber
\eea
where
\bea
\label{eq:fluxOrbits2}
& &  {h}_{ijk} = \left({H}_{ijk}  +\frac{3}{2}\, {P}_{[\underline{i}}{}^{lm} C^{(4)}_{lm\underline{jk}]}\right)~, \, \, \, \, \, {f}_{ijk}=\left({F}_{ijk}  +\frac{3}{2}\, {Q}_{[\underline{i}}{}^{lm} C^{(4)}_{lm\underline{jk}]}\,\right). 
\eea
\end{subequations}
The importance/relevance of these flux combinations will be clearer as we proceed across the various sections of this article. By using these new flux orbits which generalizes the results of \cite{Blumenhagen:2013hva, Font:2008vd}, one can rewrite the old-flux squared terms (like $H^2, F^2$ etc.) into a set of new-flux squared terms (like ${\cal H}^2, {\cal F}^2$ etc.) in a useful manner. A close inspection of the full F-term scalar potential make it possible to rearrange the various terms into the following interesting pieces,
\begin{subequations}
\bea
\label{eq:potentialSdualEin}
& & \hskip-1.5cm {\bf V_{F}} = {\bf V_{{\cal H}{\cal H}}}+ {\bf V_{{\cal F}{\cal F}}} +  {\bf V_{{\cal Q}{\cal Q}}} + {\bf V_{{\cal P}{\cal P}}} + {\bf V_{{\cal H}{\cal Q}}} + {\bf V_{{\cal F}{\cal P}}}+ {\bf V_{{\cal Q}{\cal P}}} \\
& & \hskip+1.5cm + {\bf V_{{\cal H}{\cal F}}}+ {\bf V_{{\cal F}{\cal Q}}}+ {\bf V_{{\cal H}{\cal P}}} + ....... \nonumber
\eea
where dots denote a collection of terms which could not be rearranged in new flux combinations, however such terms are precisely canceled by using the Bianchi identities which we will elaborate on later. The explicit expressions of various pieces in eqn. (\ref{eq:potentialSdualEin}) are given as under,
\bea
\label{eq:detailedV}
& & \hskip-2.5cm {\bf V_{{\cal H}{\cal H}}}= \frac{s}{4\, {\cal V}_E} \, \biggl[\frac{1}{3!}\ov{\cal H}_{ijk}\,  \ov{\cal H}_{i'j'k'}\, g_E^{ii'}\, g_E^{jj'} g_E^{kk'}\biggr]\, \nonumber\\
& & \hskip-2.5cm {\bf V_{{\cal F}{\cal F}}}= \frac{1}{4 \, s \, {\cal V}_E}\biggl[\frac{1}{3!}\ov{\cal F}_{ijk}\,  \ov{\cal F}_{i'j'k'}\, g_E^{ii'} \, g_E^{jj'} g_E^{kk'}\biggr] \nonumber\\
& & \hskip-2.5cm {\bf V_{{\cal Q}{\cal Q}}}= \frac{1}{4\, s \, {\cal V}_E} \, \biggl[3 \times \left(\frac{1}{3!}\, \ov{\cal Q}_k{}^{ij}\, \ov{\cal Q}_{k'}{}^{i'j'}\,g^E_{ii'} g^E_{jj'} g_E^{kk'} \right) + \, 2 \times \left(\frac{1}{2!}\ov{\cal Q}_m{}^{ni}\, \ov{\cal Q}_{n}{}^{mi'}\, g^E_{ii'}\right) \biggr]\nonumber\\
& & \hskip-2.5cm {\bf V_{{\cal P}{\cal P}}}= \frac{s}{4\, {\cal V}_E} \, \biggl[3 \times \left(\frac{1}{3!}\, \ov{\cal P}_k{}^{ij}\, \ov{\cal P}_{k'}{}^{i'j'}\,g^E_{ii'} g^E_{jj'} g_E^{kk'} \right) + \, 2 \times \left(\frac{1}{2!}\ov{\cal P}_m{}^{ni}\, \ov{\cal P}_{n}{}^{mi'}\, g^E_{ii'}\right) \biggr]\nonumber\\
& & \hskip-2.5cm {\bf V_{{\cal H}{\cal Q}}}= \frac{1}{4\, {\cal V}_E} \, \biggl[{\bf (-2)} \times \left(\frac{1}{2!} \ov{\cal H}_{mni} \, \ov{\cal Q}_{i'}{}^{mn}\, g_E^{ii'}\right)\biggr] \\
& & \hskip-2.5cm {\bf V_{{\cal F}{\cal P}}}= \frac{1}{4 \, {\cal V}_E} \, \biggl[{\bf (+2)} \times \left(\frac{1}{2!} \ov{\cal F}_{mni} \, \ov{\cal P}_{i'}{}^{mn}\, g_E^{ii'}\right)\biggr]\nonumber\\
& & \hskip-2.5cm {\bf V_{{\cal Q}{\cal P}}}= \frac{1}{4 \, {\cal V}_E}\, \biggl[{\bf (+2)} \times \left(\frac{1}{3!} \, \, (3\, \ov{\cal P}_{n}{}^{l'm'} \, g^E_{l' l} \, g^E_{m' m})\right)\,\, {\cal E}_E^{ijklmn} \, \, \left(\frac{1}{3!} \, (3 \, \ov{\cal Q}_{k}{}^{i'j'}\, g^E_{i'i} g^E_{j' j}) \,  \right) \biggr] \nonumber\\
& & \hskip-1.5cm \equiv  \frac{1}{4\, {\cal V}_E}\, \biggl[{\bf (+2)} \times \left(\frac{1}{2!} \, \, (\ov{\cal P}_{k'}{}^{i j} \, g_E^{k' k})\right)\, \, {\cal E}^E_{ijklmn} \, \, \left(\frac{1}{2!} (\ov{\cal Q}_{n'}{}^{l,m}\, g_E^{n'n}) \,  \right) \biggr] \nonumber\\
& & \hskip-2.5cm {\bf V_{{\cal H}{\cal F}}}=  \frac{1}{4 \, {\cal V}_E} \biggl[{\bf (+2)} \times \left(\frac{1}{3!} \, \times\, \frac{1}{3!} \, \, \ov{\cal H}_{ijk} \,\, {\cal E}_E^{ijklmn} \, \, \ov{\cal F}_{lmn}\right) \biggr] \nonumber\\
& & \hskip-2.5cm {\bf V_{{\cal F}{\cal Q}}}= \frac{1}{4\, s \, {\cal V}_E} \, \, \Big[{\bf (+2)} \times \biggl( \frac{1}{2!} \, \times\, \frac{1}{2!} \, \ov {\cal Q}_{i}{}^{j'k'} \, \ov {\cal F}_{j'k'j} \, \, \, \, \tau^E_{klmn} \,\,\, {\cal E}_E^{ijklmn} \, \,  \biggr)\Big] \nonumber\\
& & \hskip-2.5cm {\bf V_{{\cal H}{\cal P}}}=  \frac{s}{4 \, {\cal V}_E} \, \, \Big[{\bf (+2)} \times \biggl( \frac{1}{2!} \, \times\, \frac{1}{2!} \, \ov {\cal P}_{i}{}^{j'k'} \, \ov {\cal H}_{j'k'j} \, \, \, \, \tau^E_{klmn} \,\,\, {\cal E}_E^{ijklmn} \, \,  \biggr)\Big]. \nonumber
\eea
\end{subequations}
In order to understand and appreciate the nice structures within the aforementioned expressions, we need to supplement the followings,
\begin{itemize}
\item{ In the rearrangement process, we have utilized some Einstein- and string-frame conversion relations given as ${\cal V}_E = s^{3/2}\, {\cal V}_s, \, g^E_{ij} =  g_{ij} \, \sqrt{s}$ and $g_E^{ij} = g^{ij}/\sqrt{s}$ which helps us in seeing the S-duality invariance manifest.}
\item{The Levi-civita tensors are defined in terms of antisymmetric Levi-civita symbols $\epsilon_{ijklmn}$ and the same are given  as: ${\cal E}^E_{ijklmn}=\sqrt{|g_{ij}|} \, \,\epsilon_{ijklmn} = {\cal V}_E \, \, \epsilon_{ijklmn}$ while ${\cal E}_E^{ijklmn}=\epsilon^{ijklmn}/\sqrt{|g_{ij}|} =\, \epsilon^{ijklmn}/{\cal V}_E $. Further, the Einstein- and string-frame Levi-civita tensors are related as: ${\cal E}_{ijklmn} = s^{-3/2} \, {\cal E}^E_{ijklmn}$ and ${\cal E}^{ijklmn} = s^{3/2} \, \, {\cal E}_E^{ijklmn}$.}
\item{$\tau^E_{klmn}$ denotes the four-form components corresponding to the saxionic counterpart of $C^{(4)}$ RR axions with the only non-zero components being the four-cycle volume moduli, which are given as $\tau^E_{3456} = \tau_1, \tau^E_{1256} = \tau_2, \tau^E_{1234} = \tau_3$ in the notations developed in the earlier section (\ref{sec:orientifold}).}
\end{itemize}
Another motivation for the collection of terms in eqn. (\ref{eq:detailedV}) being written out only in terms of Einstein frame quantities is the fact that in our later analysis of investigating a No-Go about the universal-axion/dilaton mass splitting, we want all (inverse-)metric appearance to be independent of the dilaton. The more on this aspect will be clear in section \ref{sec:no-go}. Further, to reflect the involvement and difficulties while invoking the right combinations of flux-orbits as well as the scalar potential rearrangement, it is important to mention the following counting of terms in various pieces of the rearrangement given in eqs.(\ref{eq:potentialSdualEin})-(\ref{eq:detailedV}),
\bea
& & \hskip-1.5cm \# ({\bf V_{{\cal H}{\cal H}}}) = 1054, \, \, \, \# ({\bf V_{{\cal F}{\cal F}}}) = 4108, \, \, \, \# ({\bf V_{{\cal Q}{\cal Q}}}) = 1071, \, \, \, \# ({\bf V_{{\cal P}{\cal P}}}) = 288,\nonumber\\
& & \# ({\bf V_{{\cal H}{\cal Q}}}) = 450,\, \, \, \# ({\bf V_{{\cal F}{\cal P}}}) = 450,\, \, \, \# ({\bf V_{{\cal Q}{\cal P}}}) = 324, \\
& & \# ({\bf V_{{\cal H}{\cal F}}}) = 128,\, \, \, \# ({\bf V_{{\cal F}{\cal Q}}}) = 288,\, \, \, \# ({\bf V_{{\cal H}{\cal P}}}) = 72 \nonumber
\eea
In addition, there are 1968 terms which are removed by using Bianchi identities and are denoted as dots in eqn. (\ref{eq:potentialSdualEin}). All these numbers sum up to a total of 9661 which is the number of terms in the F-term scalar potential. For a complete detail of term-by-term analysis by turning-on a subset of fluxes at a time, see appendix (\ref{appendix_0}).

Now, the following important observations can be made out of the eqns.(\ref{eq:potentialSdualEin})-(\ref{eq:detailedV}) along with the new-orbit arrangements as mentioned in eqs. (\ref{eq:fluxOrbits1})-(\ref{eq:fluxOrbits2}),
\begin{itemize}
\item{Not only the full potential (\ref{eq:potentialSdualEin})} is manifestly S-duality invariant, but also the internal pieces $({\bf V_{{\cal H}{\cal H}}}+{\bf V_{{\cal F}{\cal F}}})$, $({\bf V_{{\cal P}{\cal P}}}+{\bf V_{{\cal Q}{\cal Q}}})$, $({\bf V_{{\cal H}{\cal Q}}}+{\bf V_{{\cal F}{\cal P}}})$ and $({\bf V_{{\cal F}{\cal Q}}}+{\bf V_{{\cal H}{\cal P}}})$ form S-duality invariant combinations while the remaining two pieces ${\bf V_{{\cal Q}{\cal P}}}$ and ${\bf V_{{\cal H}{\cal F}}}$ are self-dual.
 \item{In the absence of S-dual $P-$fluxes, one completely reproduces the results of \cite{Blumenhagen:2013hva}. Moreover, from eqns. (\ref{eq:fluxOrbits1}) and (\ref{eq:fluxOrbits2}), one can see that similar to the fact that inclusion of $Q$-fluxes corrects $F_3$-orbit by $(C_4 \bullet Q)$-type terms, the further inclusion of their dual P-fluxes modifies $H_3$-orbit with $(C_4 \bullet P$)-type terms.}
 \item{As well expected, the S-dual completion results in a more symmetrical NS-NS and RR-sector flux orbits as one can see that similar to a RR-sector flux $F_3$ having a correction of type ${\cal F}_{ijk}= F_{ijk} - C_0 \, H_{ijk}$ in Taylor-Vafa construction (and as ${\cal F}_{ijk} = {f}_{ijk} - C_0 \,{h}_{ijk}$ in the current generalized version), now we have a NS-NS flux receiving a similar type of correction from a RR-flux in the form as ${\cal Q}_k{}^{ij} = Q_k{}^{ij} - C_0 \, P_k{}^{ij}$.}
 \item{A relative minus sign in ${\bf V_{{\cal F}{\cal P}}}$ terms is observed as compared to those of ${\bf V_{{\cal H}{\cal Q}}}$ terms and the same is because of the definition of S-duality is as given in eq. (\ref{eq:S-duality}).}
 \item{Invoking the peculiar form of ${\bf V_{{\cal Q}{\cal P}}}$ is necessary as well as crucial for the oxidation process as it contains many terms of PQ-type in which all the six-indices are different and being so, such terms can neither be washed away by using PQ-type Bianchi identities nor by the anti-commutation constraints because all such respective constraints as given in eqn. (\ref{eq:BIs}) involve summation over one index. Further, the term ${\bf V_{{\cal Q}{\cal P}}}$ could be written in two equivalent ways due to the following identity,
\bea
& & \hskip-1cm \epsilon^{ijklmn} g_{i i'} g_{j j'} g_{k k'} g_{l l'} g_{m m'} g_{n n'} = |det(g_{ij})| \, \epsilon_{i'j'k'l'm'n'}
\eea}
 \item{The last three terms, namely ${\bf V_{{\cal H}{\cal F}}}$, ${\bf V_{{\cal F}{\cal Q}}}$ and ${\bf V_{{\cal H}{\cal P}}}$, are topological in nature, and so can be anticipated to be related to the contributions coming from various local sources such as brane/orientifold-tadpoles as we will elaborate now.}
\end{itemize}

\subsubsection*{Details of contributions from brane- and orientifold-sources}
In order to have the total scalar potential, the F-term contributions have to be supplemented with the D-term contributions subject to certain constraints coming from Bianchi identities (\ref{eq:BIs}) as described in detail in the appendix \ref{appendix_0}. Now as seen from the form of K\"ahler- and super-potentials, the eqn. (\ref{eq:S-duality1}) ensures that the F-term contribution is invariant under S-duality. Therefore in order to have an overall S-duality invariance, the D-terms should also be invariant and the same demands using generalized flux orbits instead of the normal ones as we will see in a moment. Further, as the pieces ${\bf V_{{\cal H}{\cal F}}}$, ${\bf V_{{\cal F}{\cal Q}}}$ and ${\bf V_{{\cal H}{\cal P}}}$ do not involve the metric unlike the rest of the terms in (of eqn. (\ref{eq:detailedV})) and happens to be topological in nature.  Moreover, the combination $({\bf V_{{\cal H}{\cal F}}}+{\bf V_{{\cal F}{\cal Q}}}+{\bf V_{{\cal H}{\cal P}}})$ is indeed S-duality invariant. Therefore, the same should be (a part of) the contributions to be compensated by imposing RR Bianchi identities or via adding the respective contribution from the various local sources. The well anticipated contributions needed from brane- and orientifold- sources to cancel the topological pieces of eqn. (\ref{eq:detailedV}) can be considered as,
\bea
\label{eq:D-term}
{\bf V_{D}} = - {\bf V_{{\cal H}{\cal F}}}- {\bf V_{{\cal F}{\cal Q}}}  - {\bf V_{{\cal H}{\cal P}}},
\eea
Here, it should be noted that $\bf{V_D}$ which is defined in terms of generalized flux combinations has a structure which is more than mere flux contributions and also contain the standard brane/orientiofld contributions coming from various local sources. Generically speaking, this ${\bf V_{D}}$ contains pieces from $D3$-brane, $O3$-plane as well as from  all the 7-branes $(D7, NS7_i, I7_i)$ as we will see in the next section where the motivation for this collection written in terms of generalized flux-combinations would be clearer for oxidation purpose. Further, it should be noted that ${\bf V_{D}}$ contributions have some pieces which can be nullified by using certain Bianchi identities given in eqn. (\ref{eq:BIs}). For example, the piece $(-{\bf V_{{\cal H}{\cal P}}})$ will have certain terms of $HP$- and $PP$-types, and the later ones can be entirely washed away by using some of the $PP$-type Bianchi identities. To be precise, out of 72 terms of $(-{\bf V_{{\cal H}{\cal P}}})$, 48 are washed away while 24 terms survive.

Now let us verify that the contributions, given in eqn. (\ref{eq:D-term}) which we also needed to compensate the topological pieces of eqn. (\ref{eq:detailedV}), indeed contain the generalized versions of D3/D7 tadpole-terms given in \cite{Blumenhagen:2013hva} with the inclusion of P-fluxes. For example, subject to applying the non-trivial Bianchi identities (\ref{eq:BIs}), switching off the P-flux recovers the following D3-tadpole terms \cite{Taylor:1999ii,Blumenhagen:2003vr},
\bea
\label{eq:VD3a}
V_{D3}^{HF} = - 2 \times \frac{1}{4 \, {\cal V}_E^2} \, \Big[ 20\, \ov{H}_{[\underline{123}}
                  \ov F_{\underline{456}]} \Big] \in (- {\bf V_{{\cal H}{\cal F}}}),
\eea
in addition to the following D7-tadpole terms (of \cite{Blumenhagen:2013hva}) given as under,
\bea
\label{eq:VD7a}
& & \hskip-1.3cm V_{D7}^{QF} = -2 \times \frac{1}{4\, s \, {\cal V}_E^2} \, \,
\Big[\ov Q_{[\underline{1}}{}^{jk} \, \ov F_{jk\underline{2}]} \, \tau_1+ \ov
Q_{[\underline{3}}{}^{jk} \, \ov F_{jk\underline{4}]}\, \tau_2+ \ov
Q_{[\underline{5}}{}^{jk} \, \ov F_{jk\underline{6}]} \, \tau_3 \Big] \in (- {\bf V_{{\cal F}{\cal Q}}}).
\eea
Now let us apply the reverse logic to motivate that in order to have S-duality invariance in the D-term contributions ($V_{D3}^{HF} + V_{D7}^{QF}$) of \cite{Blumenhagen:2013hva}, the use of generalized flux orbits is quite natural and necessary. For this purpose, consider the D7-tadpole terms $V_{D7}^{QF}$ as given in eqn. (\ref{eq:VD7a}) and invoke the terms needed for modular completion under transformations in eqn. (\ref{eq:S-duality}). Now as $1/s\longrightarrow (C_0^2 + s^2)/s$ with a S-dual of $V_{D7}^{QF}$  being of $(H P)$-type, and having in mind that $V_{D3}^{HF}$ is self-dual, one would need (at least) the following piece for a modular completion of D-term contributions,
\bea
\label{eq:VD7b}
& & \hskip-1.3cm V_{D7}^{PH} = - 2 \times \frac{\left(C_0^2+s^2\right)}{4\, s \, {\cal V}_E^2} \, \,
\Big[\ov P_{[\underline{1}}{}^{jk} \, \ov H_{jk\underline{2}]} \, \tau_1+ \ov
P_{[\underline{3}}{}^{jk} \, \ov H_{jk\underline{4}]}\, \tau_2+ \ov
P_{[\underline{5}}{}^{jk} \, \ov H_{jk\underline{6}]} \, \tau_3 \Big]
\eea
One should note that the additional piece with $C_0^2/s$ coefficient gets naturally absorbed into $(-{\bf V_{{\cal F}{\cal Q}}})$ when generalized version of fluxes ${\cal F}, {\cal H}, {\cal Q}$ and ${\cal P}$ fluxes are considered. 
Thus using generalized flux combinations rearranges the terms appropriately taking care of modular completion.

Another reason which indicates the need of our generalized flux orbits (\ref{eq:fluxOrbits1}-\ref{eq:fluxOrbits2}) essential is the fact that, the 128 terms of cross-piece $(-{\bf V_{{\cal H}{\cal F}}})$ is reduced to 32 terms and 96 terms are removed via (HQ-FP) and PP-type Binachi identities. In addition to $V_{D3}^{HF}$ which consists of 8 terms of HF-type as mentioned in eqn. (\ref{eq:VD3a}), it also results in 24 more terms of $(P^{ij}_{k} Q^{lm}_{n} \epsilon_{ijklmn})$-type which (being topological) are different from those sitting inside ${\bf V_{{\cal Q}{\cal P}}}$. Noting that neither of the QP-type Bianchi identities nor the additional anti-commutative relation in eqn. (\ref{eq:BIs}) correspond to such PQ-terms because such constraints have at least one index of QP-term being summer over, 
one should find a way to accommodate such PQ-type terms in the full picture. Interestingly, considering the generalized flux-combinations automatically does it via $(-{\bf V_{{\cal H}{\cal F}}})$, and thus resulting in no need for supplementing such strange topological terms of ${\cal Q}{\cal P}$-type.

Although, there are some more interesting aspects based on S-duality transformation of eight-form RR potential $C^{(8)}$ appearing as a triplet of eight-forms being related to produce a D-term of $({H}{ Q} + { F}{ P})$-type, however we postpone this issue to the next section, where we will discuss all the (oxidized) ten dimensional aspects.

Thus, finally following all these taxonomy of terms and taking care of contributions from the various local sources, we reach a nicely structured form of the full scalar potential given as,
\bea
\label{eq:Fullpotential}
& & \hskip-1.5cm {\bf V_{Full}} = {\bf V_{F}} + {\bf V_{D}} = {\bf V_{{\cal H}{\cal H}}}+ {\bf V_{{\cal F}{\cal F}}} +  {\bf V_{{\cal Q}{\cal Q}}} + {\bf V_{{\cal P}{\cal P}}} + {\bf V_{{\cal H}{\cal Q}}} + {\bf V_{{\cal F}{\cal P}}}+ {\bf V_{{\cal Q}{\cal P}}}
\eea
Now with this much ingredient in hand we are in a position to conjecture a modular completed version of the dimensional oxidation proposed in \cite{Blumenhagen:2013hva}.

\section{S-dual non-geometric type-IIB action: Dimensional oxidation to 10D}
\label{sec:S-dual}
With the analysis done in the previous section, a close inspection of the resulting full scalar potential, ${\bf V_{Full}} = {\bf V_{F}} + {\bf V_{D}}$ obtained as a sum of F-terms and local source contributions, reveals that all those terms can be recovered (up to satisfying a set of Bianchi identities) via a dimensional reduction from a set of generalized kinetic terms in a ten-dimensional action which, in string frame, is given as,

\begin{subequations}
\bea
\label{eq:oxiaction}
      & & \hskip-1.5cm  S={1\over 2}\int  d^{10} x\,  \sqrt{-g} \, \, \Big( {\bf {\cal L}_{\cal H \cal H}}+{\bf {\cal L}_{\cal F \cal F}}  +{\bf {\cal L}_{\cal Q \cal Q}}+{\bf {\cal L}_{\cal P \cal P}} + {\bf {\cal L}_{{\cal H}{\cal Q}}} + {\bf {\cal L}_{{\cal F}{\cal P}}} + + {\bf {\cal L}_{{\cal Q}{\cal P}}} \Big)
\eea
where
\bea
\label{eq:detailedV0}
& & {\bf {\cal L}_{\cal H \cal H}}= -{e^{-2\phi}\over 2} \, \biggl[\frac{1}{3!}\ov{\cal H}_{ijk}\,  \ov{\cal H}_{i'j'k'}\, g^{ii'}\, g^{jj'} g^{kk'}\biggr]\, \nonumber\\
& & {\bf {\cal L}_{\cal F \cal F}}= -{{1}\over 2} \, \biggl[\frac{1}{3!}\ov{\cal F}_{ijk}\,  \ov{\cal F}_{i'j'k'}\, g^{ii'} \, g^{jj'} g^{kk'}\biggr] \\
& & {\bf {\cal L}_{\cal Q \cal Q}}= -{e^{-2\phi}\over 2} \, \biggl[3 \times \left(\frac{1}{3!}\, \ov{\cal Q}_k{}^{ij}\, \ov{\cal Q}_{k'}{}^{i'j'}\,g_{ii'} g_{jj'} g^{kk'} \right)+ \, 2 \times \left(\frac{1}{2!}\ov{\cal Q}_m{}^{ni}\, \ov{\cal Q}_{n}{}^{mi'}\, g_{ii'}\right) \biggr]\nonumber\\
& & {\bf {\cal L}_{\cal P \cal P}}= -{e^{-4\phi}\over 2} \, \biggl[3 \times \left(\frac{1}{3!}\, \ov{\cal P}_k{}^{ij}\, \ov{\cal P}_{k'}{}^{i'j'}\,g_{ii'} g_{jj'} g^{kk'} \right)+ \, 2 \times \left(\frac{1}{2!}\ov{\cal P}_m{}^{ni}\, \ov{\cal P}_{n}{}^{mi'}\, g_{ii'}\right) \biggr]\nonumber\\
& & {\bf {\cal L}_{{\cal H}{\cal Q}}}= -{e^{-2\phi}\over 2} \, \biggl[{\bf (-2)} \times \left(\frac{1}{2!} \ov{\cal H}_{mni} \, \ov{\cal Q}_{i'}{}^{mn}\, g^{ii'}\right)\biggr]\nonumber\\
& & {\bf {\cal L}_{{\cal F}{\cal P}}}= -{e^{-2\phi}\over 2} \, \biggl[{\bf (+2)} \times \left(\frac{1}{2!} \ov{\cal F}_{mni} \, \ov{\cal P}_{i'}{}^{mn}\, g^{ii'}\right)\biggr]. \nonumber\\
& & {\bf {\cal L}_{{\cal Q}{\cal P}}}= -{e^{-3\phi}\over 2} \, \biggl[{\bf (+2)} \times \left(\frac{1}{2!} \, \, (\ov{\cal P}_{k'}{}^{i j} \, g^{k' k})\right)\, {\cal E}_{lmnijk} \, \left(\frac{1}{2!} \, \, (\ov{\cal Q}_{n'}{}^{l,m}\, g^{n'n}) \,  \right)\biggr]. \nonumber
\eea
\end{subequations}
This modular completed oxidation generalizes the results of \cite{Blumenhagen:2013hva}. Here, the new flux-orbits are the same as defined earlier in eqns. (\ref{eq:fluxOrbits1}-\ref{eq:fluxOrbits2}) while the (inverse-)metric components are written in their respective  string frame expressions using ${\cal V}_E = s^{3/2}\, {\cal V}_s, \, g^E_{ij} =  g_{ij} \, \sqrt{s}$ and $g_E^{ij} =  g^{ij}/\sqrt{s} \,$. Recall that string frame Levi-civita tensor is related to its Einstein frame expression as ${\cal E}_{lmnijk} = s^{-3/2} \, {\cal E}^E_{lmnijk}$. The presence of Levi-civita tensor in ${\bf {\cal L}_{{\cal Q}{\cal P}}}$ is quite anticipated for the invariance of the same as under S-duality one has $\{Q \rightarrow -P, P\rightarrow Q\}$.

Further, for capturing the correct coefficients of the respective flux-squared quantities such as $|{\cal H}|^2, |{\cal F}|^2$ etc. to those of previous section via dimensional reduction of the 10D action proposed, one has to use the ten-dimensional metric given in eqn. (\ref{eq:gMN}) as the following,
\bea
& & \hskip-1.0cm \int  d^{10} x\,  \sqrt{-g} \, (....) \simeq \int  d^{4} x\, \sqrt{-g_{\mu\nu}} \left(\frac{1}{s^{4} \, {\cal V}_s^2}\right) \times \left(\int  d^{6} x\,  \sqrt{-g_{mn}}\right) \, \times (..........) ~~ \\
& & \hskip2.2cm \simeq \int  d^{4} x\, \sqrt{-g_{\mu\nu}} \times \left(\frac{1}{s^{4} \, {\cal V}_s}\right) \times (..........). \nonumber
\eea
as $\int  d^{6} x\,  \sqrt{-g_{mn}} \equiv {\cal V}_s$ gives the string-frame 6D volume. As we can see now, the S-duality invariance in the oxidized ten-dimensional action written in string frame is not explicitly manifest as opposed to the analysis of previous section in which we kept the expressions in terms of Einstein frame quantities. In order to see the full S-duality invariance of the 10D action (\ref{eq:oxiaction}), one has to take care of transformation of the integral measure as well.

\subsubsection*{Comments on local-source contributions relating to CS-action in 10D}
After proposing the oxidized ten-dimensional kinetic terms, now let us also focus on the contributions ${\bf V_D}$, given in eqn. (\ref{eq:D-term}), which could be thought of being related to the ten-dimensional Chern-Simons terms of the following $SL(2,\IZ)$-invariant types \cite{Aldazabal:2006up, Aldazabal:2008zza, Font:2008vd, Guarino:2008ik}\footnote{Here, we have a sign difference in the first and last terms involving $C^{(4)}$ and  $\tilde C'^{(8)}$, as compared to those in \cite{Aldazabal:2006up, Aldazabal:2008zza}. This is because of the presence of a relative minus sign in $C_4$ (and $C_0$ also) while defining the chiral variables $T$ (and $S$)  as compared to their respective definitions in \cite{Aldazabal:2006up, Aldazabal:2008zza}.},
\bea
\label{eq:CS1}
& & \hskip-1.2cm S_{CS} \sim - \int C^{(4)} \wedge {F} \wedge {H}  \\
& & - \int C^{(8)} \wedge {Q}\bullet {F}\, + \int \tilde{C}^{(8)} \wedge {P}\bullet {H} \, - \int C'^{(8)} \wedge ({Q} \bullet {H} + {P} \bullet {F}) \nonumber
\eea
The first line is related to $D3/O3$-tadpoles while the second line corresponds to various 7-brane tadpoles. The first term is manifestly S-duality invariant as the RR four-form $C^{(4)}$ is $SL(2,\IZ)$ invariant, while for checking the S-duality invariance in the second line terms, one needs to consider the fact that the eight-form RR potential appears as an $SL(2, \IZ)$ triplet ($C^{(8)}$ , $\tilde C^{(8)}$ , $C'^{(8)}$) of eight-forms. These eight-form triplet components follow the S-dual transformation as
\bea
\label{eq:C8transform}
& & C^{(8)} \to - \tilde C^{(8)},\,\, \tilde C^{(8)} \to - C^{(8)},\,\, C'^{(8)}\to - C'^{(8)}
\eea
The first two of these transformation relations ensure the S-duality invariance between the first two terms of the  second line of eqn. (\ref{eq:CS1}) relating $D7$-brane and S-dual $NS7_i$-brane tadpoles \cite{Aldazabal:2006up}. Further, the sign change of $C'^{(8)}$ under S-duality ensure the survival of S-duality odd combination of fluxes $({Q} \bullet {H} + {P} \bullet {F})$ which results in the so-called $I_{7_i}$-brane tadpoles. Further as the eight-form potentials ($C^{(8)}$ , $\tilde C^{(8)}$ , $C'^{(8)}$) correspond to the dual of axion-dilaton $S$, therefore there should be some way to reduce the same into two propagating degrees of freedoms, and we will see it happening precisely while using our new flux-orbits.

Now, let us explicitly investigate the origin of our D-brane tadpoles given in eqn. (\ref{eq:D-term}) through the respective ten-dimensional Chern-Simons' action, and see how those could get related to eqn. (\ref{eq:CS1}). For this purpose, let us reconsider the expressions D-brane tadpoles being the following pieces written from eqn. (\ref{eq:D-term}) as under,
\bea
& &  \hskip-1.0cm {\bf V_D} = V_1 + V_2; \\
& & V_{1} = -\frac{1}{2 \, {\cal V}_E} \biggl[ \left(\frac{1}{3!} \, \times\, \frac{1}{3!} \, \, \ov{\cal H}_{ijk} \,\, {\cal E}_E^{ijklmn} \, \, \ov{\cal F}_{lmn}\right) \biggr] \nonumber\\
& & V_{2} = - \frac{1}{2\, s \, {\cal V}_E} \, \, \Big[ \biggl( \frac{1}{2!} \, \times\, \frac{1}{2!} \, \ov {\cal Q}_{i}{}^{j'k'} \, \ov {\cal F}_{j'k'j} \, \, \, \, \tau^E_{klmn} \,\,\, {\cal E}_E^{ijklmn} \, \,  \biggr)\Big] \nonumber\\
& & \hskip 1cm   - \frac{s}{2 \, {\cal V}_E} \, \, \Big[\biggl( \frac{1}{2!} \, \times\, \frac{1}{2!} \, \ov {\cal P}_{i}{}^{j'k'} \, \ov {\cal H}_{j'k'j} \, \, \, \, \tau^E_{klmn} \,\,\, {\cal E}_E^{ijklmn} \, \,  \biggr)\Big] \nonumber
\eea
At a first glance, it appears that terms of ${\bf V_D}$ can be trivially related to all the piece of CS action in eqn (\ref{eq:CS1}) except the last terms with a piece $({Q} \bullet {H} + {P} \bullet {F})$. However, one should recall that ${\bf V_D}$ is written out in terms of generalized flux combinations while CS-terms in eqn. (\ref{eq:CS1}) are written using normal fluxes. Let us make some more taxonomy of the respective terms. Writing back these expressions in terms of older fluxes by using our new-flux orbit definitions in eqn. (\ref{eq:fluxOrbits1}-\ref{eq:fluxOrbits2}), we find an interesting rearrangement of terms \footnote{For explicit details related to which of the terms are nullified by Bianchi identities, see full expressions of $V_1$ and $V_2$ in terms of non-generalized fluxes given in the appendix (\ref{sec:D-term}).},
\bea
& & \hskip-0.5cm V_{1} \equiv V_{1}^a + V_{1}^b \, ; \,  \nonumber\\
& & V_1^a  \, =\, - \, \frac{1}{2 \, {\cal V}_E} \biggl[ \left(\frac{1}{3!} \, \times\, \frac{1}{3!} \, \, \ov{H}_{ijk} \,\, {\cal E}_E^{ijklmn} \, \, \ov{F}_{lmn}\right) \biggr] \nonumber\\
& & \hskip0.0cm V_1^b = - \, \, \frac{1}{2 \, {\cal V}_E} \biggl[ \left(\frac{1}{3!} \, \times\, \frac{1}{3!} \, \, \left(\frac{3}{2}\, {P}_{[\underline{i}}{}^{l'm'} C^{(4)}_{l'm'\underline{jk}]}\right) \,\, {\cal E}_E^{ijklmn} \, \, \left(\frac{3}{2}\, {Q}_{[\underline{l}}{}^{l'm'} C^{(4)}_{l'm'\underline{mn}]}\right) \right) \biggr]  \nonumber\\
& & \hskip-0.5cm V_{2} \equiv V_{2}^a + V_{2}^b +V_2^c \, ; \,  \\
& & V_2^a =  - \, \, \frac{1}{2\, s \, {\cal V}_E} \, \, \Big[ \biggl( \frac{1}{2!} \, \times\, \frac{1}{2!} \, \ov {Q}_{i}{}^{j'k'} \, \ov {F}_{j'k'j} \, \, \, \, \tau^E_{klmn} \,\,\, {\cal E}_E^{ijklmn} \, \,  \biggr)\Big] \nonumber\\
& & \hskip0.0cm  V_2^b = - \, \, \frac{s^2+C_0^2}{2 \, s\, {\cal V}_E} \, \, \Big[\biggl( \frac{1}{2!} \, \times\, \frac{1}{2!} \, \ov {P}_{i}{}^{j'k'} \, \ov {H}_{j'k'j} \, \, \, \, \tau^E_{klmn} \,\,\, {\cal E}_E^{ijklmn} \, \,  \biggr)\Big] \nonumber\\
& &  V_2^c =  -\, \, \frac{1}{2 \,\, {\cal V}_E} \, \times \left(\frac{-C_0}{s}\right)\,  \Big[\biggl( \frac{1}{2!} \, \times\, \frac{1}{2!} \, \left(\ov {Q}_{i}{}^{j'k'} \, \ov {H}_{j'k'j}+\ov {P}_{i}{}^{j'k'} \, \ov {F}_{j'k'j}\right) \, \, \, \, \tau^E_{klmn} \,\,\, {\cal E}_E^{ijklmn} \, \,  \biggr)\Big] \nonumber
\eea
Now, we have a couple of peculiar and very interesting observations to make,
\begin{itemize}
\item{The term $V_1^a$ simply corresponds to the well-known D3-tadpoles in a setup without non-geometric fluxes, and
\bea
& & V_1^a \in - \int C^{(4)} \wedge {F} \wedge {H}
\eea. }
\item{Using $s\to s/(s^2+C_0^2)$ along with flux and eight-form transformations, it is clear that $V_2^a+V_2^b$ is S-duality invariant and
\bea
& & V_2^a+V_2^b  \in - \int C^{(8)} \wedge {Q}\bullet {F}\, + \int \tilde{C}^{(8)} \wedge {P}\bullet {H}.
\eea}
\item{The term $V_2^c$ with anti- S-dual combination $({Q} \bullet {H} + {P} \bullet {F})$ survives because the coefficient is also anti- S-dual as $C_0/s \to -\, C_0/s$ under S-duality. Thus, we are able to recover the last term in CS-action as
\bea
& & V_2^c  \in -\int C'^{(8)} \wedge ({Q} \bullet {H} + {P} \bullet {F}).
\eea}
\item{In addition to the four type of terms we discussed, if we use non-generalized flux orbits, there is an additional term in form of $V_1^b$. Note that, this piece contains some terms of  PQ-types in which all six flux-indices are different, and so such terms can neither be nullified by using any PQ- Bianchi identities nor using any of the anti-commutation constraints of QP-type. Therefore one needs to either introduce a new CS-term of type
\bea
& & V_1^b \in - \int C^{(4)} \wedge {\tilde P} \wedge {\tilde Q}
\eea
where $\tilde{P}_{lmn} = \left(\frac{3}{2}\, {P}_{[\underline{l}}{}^{l'm'} C^{(4)}_{l'm'\underline{mn}]}\right)$ and $\tilde{Q}_{lmn} = \left(\frac{3}{2}\, {Q}_{[\underline{l}}{}^{l'm'} C^{(4)}_{l'm'\underline{mn}]}\right)$,
or else one should seek for another way of absorbing those terms into the standard picture by rearranging the field strengths. In our case, it is the later one which happens to be true via using new flux-combinations.}
\end{itemize}
As we have mentioned earlier, these new observations also support the need of using our generalized flux combinations. In the new flux orbits, we not only embed all terms of $ ({Q} \bullet {H} + {P} \bullet {F})$ coupled with $C'^{(8)}$ eight-form into terms of type ${{\cal F}{\cal Q}}$ and ${{\cal H}{\cal P}}$, but also this helps in absorbing the additional strange looking PQ-type terms into ${\cal H} \wedge {\cal F}$.  One should note that using generic form of (QH-PF) Bianchi identity, which is given as $Q_{[\ov k}{}^{ij} H_{\ov l \ov m]j} - P_{[\ov k}{}^{ij} F_{\ov l \ov m]j} = 0$, will generically not allow the nullification of the respective terms of $({Q} \bullet {H} + {P} \bullet {F})$ though it can reduce the number of such terms \footnote{However, as pointed out in \cite{Aldazabal:2006up}, this combination $({\cal Q} \bullet {\cal H} + {\cal P} \bullet {\cal F})$ does not have RR character and, in particular cases, 
this term can be nullified. For example,  by using the following simplified version of Binachi identities does so,
\bea
{\cal Q}_{[\ov k}{}^{ij} H_{\ov l \ov m]j} = 0 =  P_{[\ov k}{}^{ij} F_{\ov l \ov m]j}\,\,\,\,\,
\Longrightarrow \,\,\,\, Q \bullet H + P \bullet F =0 .
\eea
Unlike this simplified case, there are examples of flux choices giving non-zero $I_{7_i}$-brane tadpoles in \cite{Guarino:2008ik}.}. Subsequently, we propose the following generalized form of the ten-dimensional Chern-Simons' action written in terms  of new flux-orbits as under,
\bea
\label{eq:CS}
& & \hskip-2cm S_{CS} \sim - \int C^{(4)} \wedge {\cal F} \wedge {\cal H}  - \int C^{(8)} \wedge {\cal Q}\bullet {\cal F}\, + \int \tilde{C}^{(8)} \wedge {\cal P}\bullet {\cal H} \, \, . \, \, 
\eea
\section{Robustness of the No-Go for universal-axion and dilaton mass splitting }
\label{sec:no-go}

The general four dimensional scalar potential with the inclusion of all four types of (non-)geometric fluxes ($H, F, Q$ and $P$), depend on all the 14 real moduli/axions, and a schematic form would be as under,
\bea
{\bf V} \equiv {\bf V}\left(s, c_0, \tau_i, \rho_i, u_i, v_i \right)  \, \, \, \, \, \, \, \forall i \in \{1, 2, 3\}.
\eea
Here, the scalar potential ${\bf V}$, as mentioned in eqs. (\ref{eq:Fullpotential}), denotes the sum of F-and D-term contributions which, can also be obtained from the dimensional reduction of 10D action proposed in eqns. (\ref{eq:oxiaction})-(\ref{eq:detailedV0}) along with the generalized flux orbits as in eqns. (\ref{eq:fluxOrbits1})-(\ref{eq:fluxOrbits2}). After collecting all the terms for dependencies of universal axion $c_0$ and the dilaton $s$, the very general scalar potential takes the following form,
\bea
\label{eq:Vpotsimple}
V = \left(\frac{a_1}{s} + a_2 + a_3 \, s \right) + \frac{a_4}{s} \, c_0 + \frac{a_5}{s} \, c_0^2
\eea
where $a_i$'s are generically some functions of various fluxes and all moduli/axions except universal axion $c_0$ and the dilaton $s$. This form of rearrangement of terms has been made to facilitate the study of a two-field dynamics. The extremization conditions for $c_0$ and $s$ are simply given as
\bea
& & \frac{\partial V}{\partial c_0} =  \frac{a_4+ 2 \, a_5 \, c_0}{s}\, , \, \, \, \frac{\partial V}{\partial s} = -\frac{a_1 + a_4 \, c_0 + a_5 \, c_0^2 - a_3 \, s^2}{s^2}\, . \,
\eea
This shows that if one wants $\frac{\partial V}{\partial c_0}=0$ without fixing $c_0$, then one needs to satisfy flux constraints $a_4 = 0 = a_5$, and subsequently $\frac{\partial V}{\partial s}  = -\frac{a_1 - a_3 \, s^2}{s^2}$. Now, the most crucial thing which happens to be true, is the fact that
\bea
a_3 = a_5
\eea
and the same implies that {\it ``the dilaton $s$ can not be fixed via $\frac{\partial V}{\partial s} = 0$ unless the universal axion $c_0$ is fixed via $\frac{\partial V}{\partial c_0} = 0$"}. Note that all the $a_i$-parameters generically depend on all the other moduli/axions except the universal axion and dilaton, nevertheless the above quoted argument holds independent of the fact whether those additional moduli or axions are stabilized or not. This is because of the fact that this argument is independent of the details of $a_i$s and follows from the extremization conditions of $c_0$ and dilaton. Moreover, it is worth to note that the condition: $a_3 = a_5$, holds irrespective of imposing the Bianchi identities or adding counter tadpole -terms. Now to support our arguments, we compare the scaler potential given in eqs. (\ref{eq:potentialSdualEin})-(\ref{eq:detailedV}) with our eqn. (\ref{eq:Vpotsimple}), and we get the following explicit expressions of $a_i$'s,
\bea
& & \hskip-0.5cm a_1 = \frac{1}{4\, {\cal V}_E} \times \frac{1}{3!}\, \left(\ov{F}_{ijk}+\frac{3}{2}\,\ov{Q}_{[\underline{i}}{}^{lm} C^{(4)}_{lm\underline{jk}]}\right) \left(\ov{F}_{i'j'k'} +\frac{3}{2}\,\ov{Q}_{[\underline{i'}}{}^{l'm'} C^{(4)}_{l'm'\underline{j'k'}]}\,\right) g_E^{ii'} \, g_E^{jj'} g_E^{kk'} \nonumber\\
& & \hskip-0.5cm \hskip0.7cm + \frac{1}{4\, {\cal V}_E} \, \biggl[3 \times \left(\frac{1}{3!}\, \ov{ Q}_k{}^{ij}\, \ov{ Q}_{k'}{}^{i'j'}\,g^E_{ii'} g^E_{jj'} g_E^{kk'} \right)+ \, 2 \times \left(\frac{1}{2!}\ov{ Q}_m{}^{ni}\, \ov{ Q}_{n}{}^{mi'}\, g^E_{ii'}\right) \biggr]\, ,\nonumber\\
& & \hskip-0.5cm a_2 = \frac{1}{4 \, {\cal V}_E} \, \biggl[\, 2 \times \left(\frac{1}{2!} \ov{F}_{mni} \, \ov{P}_{i'}{}^{mn}\, g_E^{ii'}\right)-2 \times \left(\frac{1}{2!} \ov{H}_{mni} \, \ov{Q}_{i'}{}^{mn}\, g_E^{ii'}\right) \nonumber\\
& & \hskip0.5cm + 2 \times \left(\frac{1}{2!} \, \, (\ov{P}_{k'}{}^{i j} \, g_E^{k' k})\right)\, {\cal E}^E_{lmnijk} \, \left(\frac{1}{2!} \, \, (\ov{Q}_{n'}{}^{l,m}\, g_E^{n'n}) \,  \right)\biggr] \, , \nonumber\\
& & \hskip-0.5cm a_3 = \frac{1}{4\, {\cal V}_E} \times \frac{1}{3!}\, \left(\ov{H}_{ijk}+\frac{3}{2}\,\ov{P}_{[\underline{i}}{}^{lm} C^{(4)}_{lm\underline{jk}]}\right) \left(\ov{H}_{i'j'k'} +\frac{3}{2}\,\ov{P}_{[\underline{i'}}{}^{l'm'} C^{(4)}_{l'm'\underline{j'k'}]}\,\right) g_E^{ii'} \, g_E^{jj'} g_E^{kk'} \nonumber\\
& & \hskip-0.5cm \hskip0.7cm + \frac{1}{4\, {\cal V}_E} \, \biggl[3 \times \left(\frac{1}{3!}\, \ov{ P}_k{}^{ij}\, \ov{ P}_{k'}{}^{i'j'}\,g^E_{ii'} g^E_{jj'} g_E^{kk'} \right)+ \, 2 \times \left(\frac{1}{2!}\ov{ P}_m{}^{ni}\, \ov{ P}_{n}{}^{mi'}\, g^E_{ii'}\right) \biggr] \nonumber\\
& & \nonumber\\
& & \equiv a_5  \, ,\\
&& \nonumber\\
& & \hskip-0.5cm a_4 = \frac{(- 2)}{4\, {\cal V}_E} \times \frac{1}{3!}\, \left(\ov{F}_{ijk}+\frac{3}{2}\,\ov{Q}_{[\underline{i}}{}^{lm} C^{(4)}_{lm\underline{jk}]}\right) \left(\ov{H}_{i'j'k'} +\frac{3}{2}\,\ov{P}_{[\underline{i'}}{}^{l'm'} C^{(4)}_{l'm'\underline{j'k'}]}\,\right) g_E^{ii'} \, g_E^{jj'} g_E^{kk'}\nonumber\\
& & \hskip-0.5cm \hskip0.7cm+ \frac{(- 2)}{4\, {\cal V}_E} \, \biggl[3 \times \left(\frac{1}{3!}\, \ov{P}_k{}^{ij}\, \ov{Q}_{k'}{}^{i'j'}\,g^E_{ii'} g^E_{jj'} g_E^{kk'} \right)   + \, 2 \times \left(\frac{1}{2!}\ov{P}_m{}^{ni}\, \ov{Q}_{n}{}^{mi'}\, g^E_{ii'}\right) \biggr] ~. \nonumber
\eea
Let us point out that by looking at the S-duality transformation, we observe that:
\bea
& & a_1 \leftrightarrow a_3 \equiv a_5, a_2 \rightarrow a_2, a_4 \rightarrow -a_4.
\eea
which using $s\longrightarrow s/(c_0^2 + s^2)$ and $c_0/s \longrightarrow - c_0/s$ ensure the S-duality invariance of the total potential in $\{c_0, s\}$ variables. The full potential can be written in S-dual pieces after using $a_3 = a_5$ as below,
\bea
\label{eq:Vpotsimple1}
V =  a_2 + \left(\frac{a_1}{s}  + {a_3} \, \frac{(c_0^2 + s^2)}{s} \right) + {a_4} \, \frac{c_0}{s}
\eea
The next question is whether it is possible to create a hierarchy via additional fluxes when we stabilize $c_0$ and $s$ simultaneously. To address this question will need a complete minimization analysis of the full scalar potential with 14 scalars along with an overall 64 flux components ! Although it will be a bit strong assumption to make, let us consider the parameters $a_i$'s as constants and simply investigate the dynamics of two fields, namely the universal axion and the dilaton, appearing in the same chiral multiplet $S$. Subsequently, the Hessian at one set of critical point: $\ov c_0 = -\frac{a_4}{2 \, a_5}, \ov s = \frac{\sqrt{4 a_1 \, a_5 - a_4^2}}{2 \sqrt{a_3}\, \sqrt{a_5}}$ is given as under
\bea
& & V_{c_0 c_0} = \frac{4\, \sqrt{a_3} \, a_5^{3/2}}{\sqrt{4 a_1 \, a_5 - a_4^2}}, \, V_{c_0 s} = 0 = V_{s c_0}, \,  V_{s s} = \frac{4\, \sqrt{a_5} \, a_3^{3/2}}{\sqrt{4 a_1 \, a_5 - a_4^2}},
\eea
which implies that
\bea
\frac{m_{c_0}^2}{m_s^2} = \frac{a_5}{a_3} = 1.
\eea
So, with this two-field analysis, we can anticipate that it is not possible to have mass splitting of the chiral multiplet $S = e^{-\phi} - i\, C_0$  even with the inclusion of non-geometric fluxes. However as mentioned earlier, for the complete analysis, one has to investigate the full Hessian matrix of size $14\times14$, and carefully look at the non-trivial off-diagonal entries while diagonalizing the mass-matrix. 

Thus, our investigation recovers the claim of \cite{Blumenhagen:2014nba} about the impossibility keeping the universal axion massless while stabilizing the dilaton in the simplest Taylor-Vafa construction \cite{Taylor:1999ii,Blumenhagen:2003vr} in the absence of non-geometric ($Q$, $P$) fluxes. In addition, our analysis supports for the validity of  the first part of the No-Go theorem \cite{Blumenhagen:2014nba} that {while considering a two-field dynamics, one can not have a mass splitting in universal axion and dilaton masses even with the help of S-dual pairs of non-geometric fluxes}. However, models with additional contributions to the scalar potential
may also avoid this no-go theorem.
Such corrections can involve D-brane instanton
effects to the non-perturbative superpotential, or perturbative
corrections to the K\"ahler potential for the K\"ahler moduli, which can break the no-scale structure and such effects should be studied in great detail.

\section{Conclusion}
In this article, we propose a S-duality invariant ten-dimensional supergravity action via dimensional oxidation of a four-dimensional scalar potential,  obtained by utilizing the K\"ahler- and super-potential expressions for a toroidal orientifold of type IIB superstring theory in the presence of non-geometric fluxes. In this context, we have generalized the flux orbits of \cite{Blumenhagen:2013hva} with the inclusion of RR P-flux being S-dual of the non-geometric Q-flux, and these generalized flux combinations appearing in ten-dimensional kinetic terms  are as follows,
\begin{subequations}
\bea
&& {\cal H}_{ijk} =  {h}_{ijk}~, \, \, \, \, \, {\cal Q}^{ij}_{k} = {Q}^{ij}_{k} - C_0 \, ~{P}^{ij}_{k} ~, \nonumber\\
&& {\cal F}_{ijk}= {f}_{ijk} - C_0 \, ~{h}_{ijk}~, \, \, \, \, \, {\cal P}^{ij}_{k} = {P}^{ij}_{k}~ .\nonumber
\eea
where
\bea
& & \hskip-1cm {h}_{ijk} = \left({H}_{ijk}  +\frac{3}{2}\, {P}_{[\underline{i}}{}^{lm} C^{(4)}_{lm\underline{jk}]}\right)~, ~~~{f}_{ijk}=\left({F}_{ijk}  +\frac{3}{2}\, {Q}_{[\underline{i}}{}^{lm} C^{(4)}_{lm\underline{jk}]}\,\right). \nonumber
\eea
\end{subequations}
We have motivated and exemplified the need for the use of these generalized flux-combinations in many stages; not only in nicely arranging the ten-dimensional kinetic terms out of F-term contribution of the scalar potential but also in consistently reproducing the S-dual version of the ten dimensional Chern-Simons'-terms via the D-brane tadpoles. In addition, we find that using our new flux orbits, only two propagating dofs out of the three eight-form triplet potentials ($C^{(8)}$ , $\tilde C^{(8)}$ , $C'^{(8)}$) survive which is consistent as well as desirably compatible because RR eight-form is dual to the axion-dilaton $S$.

As an application of the explicit expressions obtained, we examined the recently proposed No-Go theorem \cite{Blumenhagen:2014nba} about the impossibility of mass-splitting of axion-dilaton chiral multiplet, and investigating a two-field dynamics with fields $c_0$ and $s$ assuming that all the other moduli/axion are fixed at their minimum, we find that the No-Go result still holds with the inclusion of non-geometric Q- and its S-daul P-flux as well. However, for a final conclusion, one needs to minimize the full potential by considering the dynamics of all the 14 scalars with the presence of 64 consistent flux parameters. Further, it would be also interesting to check for the possibility alleviating the No-Go by non-perturbative effects in the presence of Non-geometric fluxes.  Although, with the present poor understanding, it is hard to make any conclusion about the influence of non-geometric fluxes through non-perturbative effects, nevertheless, something robust happening at tree level would be expected to remain intact by sub-leading corrections. It would be also crucially important to perform a very detailed moduli stabilization, and to hunt for other combination of axionic directions which could be sufficiently lighter for satisfying the inflationary requirements.

\section*{Acknowledgments}
We gratefully acknowledge very significant learning on the subject from Ralph Blumenhagen and Daniela Herschmann during a previous collaboration. We are also very thankful to the referee for her/his very useful and enlightening suggestions and queries. XG would like to thank Lara Anderson, James Gray and Seung-Joo Lee for useful discussions. The work of XG was supported in part by NSF grant PHY-1417337. PS was supported by the Compagnia di San Paolo contract ``Modern Application of String Theory'' (MAST) TO-Call3-2012-0088.

\clearpage
\appendix

\section{Counting of terms for invoking the various pieces of oxidized 10-D action}
\label{appendix_0}
Here we present a detailed analysis along with the intermediate steps taken for matching the two actions; the first one coming from K\"ahler potential and superpotential as a F-term contribution, ${\bf V_{F}}$, while the other one $V_{\rm kin}$, as given in form of various kinetic terms through expressions (\ref{eq:potentialSdualEin})-(\ref{eq:detailedV}). The later one is expected to come from the dimensional reduction of an oxidized 10-dimensional non-geometric action given in eqs.(\ref{eq:oxiaction}) and (\ref{eq:detailedV0}). For guessing the form of oxidized 10-dimensional action, the strategy has been as under,
\begin{enumerate}
\item{First, we collected all 9661 terms appearing as F-term contribution to the four-dimensional scalar potential, $V_F$, obtained by using the K\"ahler- and super-potentials given in eq. (\ref{typeIIBK}) and (\ref{typeIIBW}) respectively.}
\item{Second, we started to look for the completion of various flux orbits obtained as in eqs. (\ref{eq:fluxOrbits1})-(\ref{eq:fluxOrbits2}) such that the respective terms in $V_{\rm kin}$ are recovered in F-term contributions, $V_{F}$. This is what we called a suitable rearrangement of F-terms. For this purpose. we considered the guidelines from earlier work \cite{Villadoro:2005cu} for type IIA with geometric-flux, and \cite{Blumenhagen:2013hva} for type II theories with non-geometric Q-flux also. This step led us with a rearrangement the 7693 terms of the full F-term potential in ten pieces of the from ${\bf V_{\cal A \cal B}}$ written out in using generalized flux-orbits $\cal A, \cal B \in \{\cal H, \cal F, \cal Q, \cal P\}$. Moreover, we found that 3 pieces out of 10, namely ${\bf V_{\cal H \cal F}}$, ${\bf V_{\cal F \cal Q}}$ and ${\bf V_{\cal H \cal P}}$, are topological in nature and could be related to minus of a D-term contribution. Such topological terms are 488 in numbers (and get split as 128 + 288 + 72 respectively) which after imposing Bianchi identities further reduces into a total of 152 terms as mentioned in Table [\ref{tableTypeIIB}]. }
\item{After recovering 7693 terms (out of 9661) terms of ${\bf V_F}$, in the final step, we are then left with 1968 terms from ${\bf (V_F + V_D)-V_{\rm kin}}$. These 1968 terms are ensured to be nullified by utilizing the following types of Bianchi identities \cite{Aldazabal:2006up,Aldazabal:2008zza},
\bea
\label{eq:BIs}
& & \hskip-1.0cm 120 \#: QQ{\rm-type:}  \, \, \, \, \, \, Q_k{}^{[\ov i \, \ov j} Q_n{}^{\ov l] {k}}=0\, , \nonumber\\
& & \hskip-1.0cm 240 \#: PP{\rm-type:}  \, \, \, \, \, \, P_k{}^{[\ov i \, \ov j} P_n{}^{\ov l] {k}}=0\, , \\
& & \hskip-1.0cm 240 \#: (HQ-FP){\rm-type:}  \, \,  \, \,  \, \, Q_{[\ov k}{}^{ij} H_{\ov l \ov m]j} -  P_{[\ov k}{}^{ij} F_{\ov l \ov m]j}\, = 0 , \nonumber\\
& & \hskip-1.2cm 1368 \#: (QP){\rm-type:}  \, \,  \, \,  \, \, Q_k{}^{[\ov i \, \ov j} P_n{}^{\ov l] {k}}\, = 0\, , \,  \, P_k{}^{[\ov i \, \ov j} Q_n{}^{\ov l]{k}}=0\, , \nonumber\\
& & \hskip4.0cm    Q_p{}^{a b} P_m{}^{pc} - P_p{}^{a b} Q_m{}^{pc}=0, \nonumber
\eea}
\end{enumerate}
where the last QP-type constraints are demanded from the anti-symmetry of the commutators involved in the derivation of the various Bianchi Identities \cite{Aldazabal:2008zza}.
\begin{table}[H]
  \centering
  \begin{tabular}{|c||c|c|c||c|c|}
  \hline
  Fluxes  & ${\bf V_{F}}$  & ${\bf V_{\rm kin}}$ & ${\bf V_F-V_{\rm kin}}$ & ${\bf V_{D}} = V_{D} + BIs$ &  ${\bf (V_{F}+V_D)-V_{\rm kin}}$\\
turned-on & & &  &  & (to be removed by BIs)  \\
\hline
\hline
& & & &   & \\
 $H$ & 152 & 152 & 0  & 0=0+0  & 0\\
 $F$ &  76 & 76 & 0  & 0=0+0  & 0\\
 $Q$ &  1059 & 891 & 168  & 48=0+48  & 120\\
 $P$ & 2118 &  1782 & 336 & 96=0+96   & 240\\
& & &   & & \\
$H, F$ & 361 &353 & 8 & 8=8+0  &  0\\
$H, Q$ & 1814 &1478 & 336 & {\bf 96=0+96}   &  240\\
$H, P$ & 3068 & 2684 & 384  & 144=48+96  & 240\\
$F, Q$ &1534 & 1342 & 192  & 72=24+48 &  120\\
$F, P$ & 2797 & 2293 & 504 &  {\bf 144=0+144}   &  360\\
{$Q, P$} & {6897} & {4857} & {2040} & 312=24+288 & 1728\\
& & & &  & \\
$H, F, Q$ & 2422 & 2054 & 368 & 128=32+96  & 240 \\
$H, F, P$ & 3880 & 3320 & 560 & 200=56+144  & 360\\
$H, Q, P$ & 8450 & 6194 & 2256 & 408=72+336  & 1848\\
$F, Q, P$ &7975 & 5743 & 2232 & 384=48+336 & 1848\\
& & & &   & \\
$H, F, Q, P$ & 9661 & 7205& 2456 & 488=152+336 & 1968 \\
 \hline
\end{tabular}
\caption{Number of individual terms with presence of a particular (set of) fluxes being turned-on at a time in the scalar potential.}
\label{tableTypeIIB}
\end{table}
Here, one should note the following observations,
\begin{itemize}
 \item{While mentioning the counting of D-terms, ${\bf V_D}$, we have considered it as ${\bf V_{D}} = V_{D} + BIs$ which is such that $V_{D}$ represents only those terms which could survive after the application of various Bianchi identities given in eqn. (\ref{eq:BIs}). This analysis was needed to investigate the CS action reproducing the D-brane tadpoles.}
 \item{In the two rows with ($H, Q$)-only and ($F,P$)-only fluxes, we find that although there are tadpoles expected from (HQ+FP)-type CS-action with $C'{(8)}$ RR-potential but while switching off a set of two fields ($H,Q$) or ($F,H$), simplifies the $(HQ-FP)$-type Bianchi identity into $HQ =0$ or $FP = 0$ case, and subsequently, no non-zero D-terms could get induced.}
 \item{As $(HQ-FP)$-type Bianchi identity is the only type which involves all the four type of fluxes, the previous argument happens to be true in case of vanishing any one of the fluxes in the combinations $(H, F, Q)$, $(H, F, Q)$, $(H, F, Q)$ and $(H, F, Q)$. However, as soon as all the four type of fluxes are generically turned-on, one gets additional terms for (HQ+FP)-type $I_{7_i}$-brane tadpoles. }
 \item{Counting in the row with only-H and only-F fluxes corresponds to standard Tayor-Vafa setup \cite{Taylor:1999ii,Blumenhagen:2003vr} while the one with H-, F- and Q-fluxes corresponds to \cite{Blumenhagen:2013hva}.}
\end{itemize}

\section{Rearrangement of D-terms for invoking the complete 10D CS action}
\label{sec:D-term}
\bea
 & & \hskip-0.75cm V_{1} \equiv  - \, \frac{1}{2 \, {\cal V}_E} \biggl[ \left(\frac{1}{3!} \, \times\, \frac{1}{3!} \, \, \ov{\cal H}_{ijk} \,\, {\cal E}_E^{ijklmn} \, \, \ov{\cal F}_{lmn}\right) \biggr]\\
 & & =\, -\frac{1}{2 \, {\cal V}_E} \left(\frac{1}{3!} \, \times\, \frac{1}{3!}\right) \, \, \biggl[ {\ov{H}_{ijk}  \, \,  {\cal E}_E^{ijklmn}\,\, \ov{F}_{lmn}} + \frac{9}{4}\,{\left( {P}_{[\underline{i}}{}^{l'm'} C^{(4)}_{l'm'\underline{jk}]}  \, \,  {\cal E}_E^{ijklmn}\, {Q}_{[\underline{l}}{}^{l'm'} C^{(4)}_{l'm'\underline{mn}]}\right)}  \nonumber\\
  & & - \frac{3}{2}\, C_0 \left( {P}_{[\underline{i}}{}^{l'm'} C^{(4)}_{l'm'\underline{jk}]}  \, \,  {\cal E}_E^{ijklmn}\,\,\cancel{ \ov{H}_{lmn}+ \ov{H}_{ijk}  \,} \,  {\cal E}_E^{ijklmn}\, \, {P}_{[\underline{l}}{}^{l'm'} C^{(4)}_{l'm'\underline{mn}]}\right)\nonumber\\
 & & -C_0 \, \cancel{{\ov{H}_{ijk}  \, \,  {\cal E}_E^{ijklmn}\,\, \ov{H}_{lmn}}} - \frac{9}{4}\, C_0 \, \cancel{\left( {P}_{[\underline{i}}{}^{l'm'} C^{(4)}_{l'm'\underline{jk}]}  \, \,  {\cal E}_E^{ijklmn}\, {P}_{[\underline{l}}{}^{l'm'} C^{(4)}_{l'm'\underline{mn}]}\right)} \nonumber\\
 & & + \frac{3}{2}\, \left( {P}_{[\underline{i}}{}^{l'm'} C^{(4)}_{l'm'\underline{jk}]}  \, \,  {\cal E}_E^{ijklmn}\,\,\cancel{ \ov{F}_{lmn}+ \ov{H}_{ijk}  \, }\,  {\cal E}_E^{ijklmn}\, \, {Q}_{[\underline{l}}{}^{l'm'} C^{(4)}_{l'm'\underline{mn}]}\right)\biggr] \nonumber
\eea
The first two cancellations are trivial mathematical ones while the last two corresponds to some parts of $PP$-type and $(QH-FP)$-type Bianchi Identities.
\bea
& & \hskip-0.75cm  V^{(i)}_{2} \equiv - \frac{1}{2\, s \, {\cal V}_E} \, \, \Big[ \biggl( \frac{1}{2!} \, \times\, \frac{1}{2!} \, \ov {\cal Q}_{i}{}^{j'k'} \, \ov {\cal F}_{j'k'j} \, \, \, \, \tau^E_{klmn} \,\,\, {\cal E}_E^{ijklmn} \, \,  \biggr)\Big] \nonumber\\
& & = - \frac{1}{8\, s \, {\cal V}_E} \, \, \biggl[\biggl(\ov {Q}_{i}{}^{j'k'} \, \ov {F}_{j'k'j} \, \, \tau^E_{klmn} \, {\cal E}_E^{ijklmn} \,\biggr) -C_0 \biggl(\ov {P}_{i}{}^{j'k'} \, \ov {F}_{j'k'j}\,  \tau^E_{klmn} \, {\cal E}_E^{ijklmn}\biggr) \\
& & \hskip1.3cm + C_0^2 \, \biggl(\ov {P}_{i}{}^{j'k'} \, \ov {H}_{j'k'j} \, \, \tau^E_{klmn} \, {\cal E}_E^{ijklmn} \biggr) \,- C_0 \biggl(\ov {Q}_{i}{}^{j'k'} \, \ov {H}_{j'k'j} \, \, \tau^E_{klmn} \, {\cal E}_E^{ijklmn} \,\biggr) \nonumber\\
& & + \frac{3}{2}\, C_0^2 \biggl(\cancel{\ov {P}_{i}{}^{j'k'} \, {P}_{[\underline{j'}}{}^{l'm'} C^{(4)}_{l'm'\underline{k'j}]} \, \, \tau^E_{klmn} \, {\cal E}_E^{ijklmn}}\biggr) + \frac{3}{2}\,\cancel{\biggl(\ov {Q}_{i}{}^{j'k'} \, {Q}_{[\underline{j'}}{}^{l'm'} C^{(4)}_{l'm'\underline{k'j}]} \, \, \tau^E_{klmn} \, {\cal E}_E^{ijklmn}\biggr)}\nonumber\\
& & - \frac{3}{2}\, C_0 \biggl(\cancel{\ov {P}_{i}{}^{j'k'} \, {Q}_{[\underline{j'}}{}^{l'm'} C^{(4)}_{l'm'\underline{k'j}]} \, \, \tau^E_{klmn} \, {\cal E}_E^{ijklmn}}\biggr) - \frac{3}{2}\,C_0\,\cancel{\biggl(\ov {Q}_{i}{}^{j'k'} \, {P}_{[\underline{j'}}{}^{l'm'} C^{(4)}_{l'm'\underline{k'j}]} \, \, \tau^E_{klmn} \, {\cal E}_E^{ijklmn}\biggr)} \biggr]\nonumber
\eea
The first two cancellations correspond to a subset of $PP$-type and $QQ$-type while the last two corresponds to some parts of $QP+PQ$-type Bianchi identities. Notice the presence of terms with coefficient $C_0^2/s$ to make $V^{(ii)}_{2}$, which is given below, S-duality invariant. The two pieces with a coefficient $(-C_0/s)$ is expected to correspond to the $I_{7_i}$-brane tadpoles coming from a (HQ+FP)-combination with anti- S-dual eight-form potential $C'^{(8)}$.
\bea
& & \hskip-0.75cm V^{(ii)}_{2} \equiv - \frac{s}{2 \, {\cal V}_E} \, \, \Big[\biggl( \frac{1}{2!} \, \times\, \frac{1}{2!} \, \ov {\cal P}_{i}{}^{j'k'} \, \ov {\cal H}_{j'k'j} \, \, \, \, \tau^E_{klmn} \,\,\, {\cal E}_E^{ijklmn} \, \,  \biggr)\Big]\\
& & = - \frac{s}{8 \, {\cal V}_E} \Big[\biggl(\ov { P}_{i}{}^{j'k'} \, \ov { H}_{j'k'j} \,  \tau^E_{klmn} \, {\cal E}_E^{ijklmn} \biggr) + \cancel{\biggl(\ov {\cal P}_{i}{}^{j'k'} \,  \left(\frac{3}{2}\, {P}_{[\underline{j'}}{}^{l'm'} C^{(4)}_{l'm'\underline{k'j}]}\right) \, \, \tau^E_{klmn} \, {\cal E}_E^{ijklmn} \biggr)}\Big] \nonumber
\eea
The last cancellation piece corresponds to a part of $PP$-type Bianchi identities.

\newpage
\bibliographystyle{utphys}
\bibliography{reference}

\end{document}